\documentclass[conference,a4paper]{IEEEtran}

\IEEEoverridecommandlockouts
\usepackage{cite}
\usepackage{amsmath,amssymb,amsfonts}
\usepackage{algorithmic}
\usepackage{graphicx}
\usepackage{balance}
\usepackage{eurosym}
\usepackage{cases}
\newcommand{\figref}[1]{Fig.~\ref{#1}}
\usepackage{textcomp}
\usepackage{hyperref}
\hypersetup{colorlinks=true, unicode=true, linkcolor=[rgb]{0.10,0.05,0.67}, citecolor=[rgb]{0.10,0.05,0.67}, filecolor=[rgb]{0.10,0.05,0.67}, urlcolor=[rgb]{0.10,0.05,0.67}}
\usepackage{dblfloatfix}
\usepackage{cases}
\usepackage{nomencl}
\usepackage{amssymb}
\usepackage{flushend}
\usepackage{etoolbox}
\usepackage{dirtytalk}
\usepackage{optidef} 
\usepackage{tabulary}
\usepackage{acronym}
\usepackage{colortbl}
\usepackage{soul}
\usepackage[normalem]{ulem}
\usepackage{amsmath}
\newcommand{\stkout}[1]{\ifmmode\text{\sout{\ensuremath{#1}}}\else\sout{#1}\fi}

\usepackage[symbol]{footmisc}
 \newcommand\footnoteref[1]{\protected@xdef\@thefnmark{\ref{#1}}\@footnotemark}

\renewcommand\nomgroup[1]{%
  \item[\bfseries
  \ifstrequal{#1}{A}{Sets}{%
  \ifstrequal{#1}{B}{Parameters}{%
  \ifstrequal{#1}{C}{Variables}{}}}%
]}

\renewcommand{\figurename}{Fig. }

\usepackage{xcolor}

\def\BibTeX{{\rm B\kern-.05em{\sc i\kern-.025em b}\kern-.08em
    T\kern-.1667em\lower.7ex\hbox{E}\kern-.125emX}}
\begin{document}

\title{Impact of Lead Time on Aggregate EV Flexibility for Congestion Management Services
\thanks{The research was supported by the ROBUST (MOOI32014) project, which received funding from the MOOI subsidy programme by the Netherlands Ministry of Economic Affairs and Climate Policy and the Ministry of the Interior and Kingdom Relations, executed by the Netherlands Enterprise Agency.}

}
\author{
\IEEEauthorblockN{Nanda Kishor Panda, Peter Palensky, Simon H.~Tindemans}
\IEEEauthorblockA{Department of Electrical Sustainable Energy\\
Delft University of Technology\\
Delft, The Netherlands}
\{n.k.panda, p.palensky, s.h.tindemans\}@tudelft.nl
}

\maketitle

\begin{abstract}
 Increased electrification of energy end-usage can lead to network congestion during periods of high consumption. Flexibility of loads, such as aggregate smart charging of Electric Vehicles (EVs), is increasingly leveraged to manage grid congestion through various market-based mechanisms. Under such an arrangement, this paper quantifies the effect of lead time on the aggregate flexibility of EV fleets. Simulations using real-world charging transactions spanning over different categories of charging stations are performed for two flexibility products (redispatch and capacity limitations) when offered along with different business-as-usual (BAU) schedules. Results show that the variation of tradable flexibility depends mainly on the BAU schedules, the duration of the requested flexibility, and its start time. Further, the implication of these flexibility products on the average energy costs and emissions is also studied for different cases. Simulations show that bidirectional (V2G) charging outperforms unidirectional smart charging in all cases. 

\end{abstract}

\acrodef{EV}{Electric Vehicle}
\acrodef{BAU}{Business as usual}
\acrodef{CPO}{Charging Point Operator}
\acrodef{CS}{Charging Station}
\acrodef{CP}{Charging Point}
\acrodef{DSM}{Demand Side Management}
\acrodef{DSO}{Distribution System Operator}
\acrodef{MEF}{Marginal Emission Factor}
\acrodef{V2G}{Vehicle-to-grid}
\acrodef{OCPI}{Open Charge Point Interface}
\acrodef{TSO}{Transmission System Operator}
\acrodef{RL}{Reinforcement Learning}
\acrodef{EVSE}{Electric Vehicle Supply Equipment}
\section{Introduction}
The widespread electrification of sectors such as heating, transportation, and industries is increasingly straining electricity networks. Especially with the large-scale adoption of \acp{EV}, power networks face challenges in safely accommodating their aggregate charging needs. This has resulted in congestion, where parts of the network (such as cables, transformers, etc.) risk overloading during high-demand times. The resulting congestion in power networks is already prominent in North America \cite{li2024impact} and mainland Europe, including the Netherlands \cite{capaciteitskaart_netbeheer_nederland}. It is only a matter of time before these issues become prevalent worldwide. One reason for the congestion is that the power networks were originally not designed for such a degree of electrification, and the simultaneity of high-power loads further exacerbates it. As a result, the grid needs to be adapted and reinforced to accommodate the growing loads. However, these need substantial investments and a longer time, making the pursuit of short-term alternatives increasingly critical~\cite{resch2021techno}.\par

Different short-term approaches, such as hard curtailments on charging power during anticipated congestion or refusal of new network connections until the congestion situation improves, are some solutions that network operators in countries like the Netherlands and Germany adopt~\cite{hennig2024risk}. However, such solutions manage congestion at the cost of end-user satisfaction, which is diminished due to lower charging speed, cost of charging, or restricted autonomy by these mandates. Mandate-based (direct-control) solutions offer limited flexibility to end-users and aggregators, potentially slowing the adoption of \acp{EV}. To address this, \acp{DSO} and policymakers increasingly favour incentive-based congestion management, which encourages voluntary participation from aggregators and end-users~\cite{venegas2021active}. The voluntary participation for \acp{EV} is possible without compromising on operational constraints due to \ac{EV}'s flexibility.\par

\acp{EV} inherently offer charging and storage (in the case of \ac{V2G} through discharging) flexibility. In other words, the time required to fully charge an \ac{EV} is typically disproportionate to its average connection duration. Leveraging this, smart charging (and \ac{V2G}) offers significant potential to mitigate (or exacerbate) network congestion.  Influencing individual charging profiles through time-of-use has been used extensively in Europe, ranging from simple dynamic energy prices to network tariffs or combinations of both~\cite{li2023residential,shen2022coordination}. Price-based approaches can  help limit aggregate charging peaks; however, if not carefully designed, they can create newer peaks or penalize consumers unfairly~\cite{panda2024aggregate}.\par

Alternatively, leveraging aggregate flexibility to manage congestion has proven highly efficient, as it can mitigate the uncertainties associated with an individual \acp{EV}~\cite{panda2024quantifying}. In addition to direct control of aggregate \acp{EV}, market-based methods are one of the most cost-effective and scalable solutions to leverage this flexibility for congestion management. These offer scalability, incentivize flexible energy usage, and promote competition among participants~\cite{hennig2023congestion}.\par

Market-based congestion management services are currently being developed in several territories as alternatives. Market-based capacity management can typically be achieved through reduction options, activation before anticipated congestion, or direct trading. In the Netherlands, the energy regulatory body has defined network congestion, assigned specific responsibilities to stakeholders participating in congestion management services and described the characteristics of market products that can help alleviate congestion~\cite{TheNetherlands2022DecreeManagement}. These include on-demand capacity limitation contracts and location-based redispatch, linked to intra-day markets~\cite{panda2024quantifying}.\par

However, barriers exist to exchanging congestion management services by \acp{CPO} to the \acp{DSO}. Besides technical challenges, such as bundling connections and communicating with numerous assets, it is unclear how much flexibility a fleet of assets can reliably offer and how it depends on the service type and the advance notice given for activation. This paper investigates how lead times and length of flexibility request windows influence the magnitude of two congestion management products—redispatch and capacity limitation—across different categories of \acp{CS} and dispatch strategies. We further compare the capabilities of \ac{V2G} or bidirectional charging with unidirectional charging (also called smart charging). The congestion management products in The Netherlands are used as a reference for model development.\par

\section{Mechanism of congestion management }
\begin{figure}
    \centering
    \includegraphics[width=0.95\linewidth]{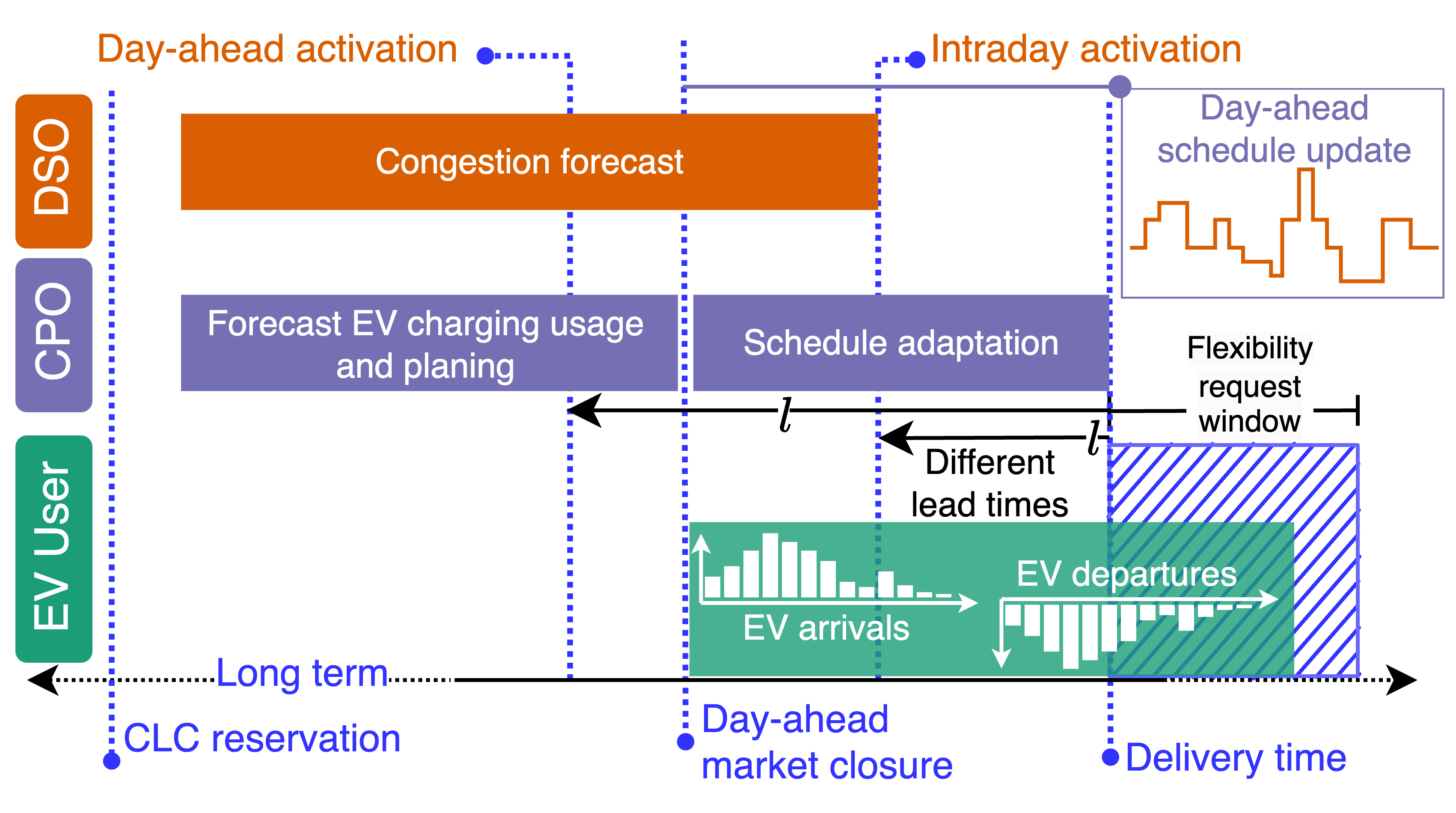}
    \caption{Schematic showcasing various market interactions alongside market
timeline for capacity limitation products.}
    \label{fig:market_timeline}
\end{figure}

\figref{fig:market_timeline} illustrates different market interactions with respect to the time horizon for the redispatch and capacity limitation products in the Netherlands. When delivering services using \acp{EV} to relieve congested distribution grids, three stakeholders (EV users, \ac{DSO} and \ac{CPO}) interact. For simplicity, the \ac{CPO} controls the charging of \acp{EV} and acts as the aggregator. The \ac{DSO} is responsible for forecasting the periods of congestion along with its duration and magnitude. Based on these forecasts, it sends activation signals to the flexibility aggregators. Based on the arrangements, activation can be done either a day ahead ($\geq$ 24 hours, or, more accurately, before clearing of the day-ahead market) or within the intraday time frame, with lead times ranging from as short as 1 hour up to the day-ahead market time frame. We define lead time as the time available to the aggregators for schedule adaption between activation and delivery of the flexibility services. This paper considers a range of 1 to 23 hours. \par

Until the aggregators are notified of activation, they schedule their \acp{EV} according to \ac{BAU} schedules. Once notified, the day-ahead schedules are adjusted to deliver the specified product effectively. Based on these interactions, the ability to trade a specific product depends on several factors such as lead time, duration of flexibility request window, charging strategy (unidirectional or \ac{V2G}), lead time, \ac{BAU} scheduling strategy and the type vehicles in the fleet.\par

Shorter lead times provide more room to efficiently utilize the network's available capacity while maximizing the \ac{CPO}'s revenue based on \ac{BAU} schedules. However, longer lead times may reduce the attractiveness of congestion management products for the \ac{DSO} as the uncertainty surrounding the occurrence of congestion increases. Conversely, shorter lead times can critically hinder the aggregators' ability to deliver these products, increasing the chances of enforced measures such as load curtailment, which could negatively impact end-user experience. The duration for which a particular product is requested also influences the magnitude of the product. As the window length increases, the magnitude keeps on diminishing.\par

Assuming that an \ac{EV} fleet has a certain energy demand to be fulfilled within a specified time frame but with scope to adjust the schedule; we consider the fleet to have \emph{potential flexibility}. This translates into \emph{feasible flexibility} once all operational and technical constraints are accounted for, including those derived from day-ahead \ac{BAU} charging schedules, and a model representation is used. The aggregator aims to maximize its revenue using congestion management products, for which it must determine the \emph{tradable flexibility} in flexibility markets - which also depends on the product requirements and lead times. Previous work \cite{panda2024quantifying} analysed the aggregate flexibility of \ac{EV} charging stations to deliver redispatch and offer aggregate capacity limitations (without sacrificing the quality of service). This was done for three distinct categories of \acp{CS} (residential, commercial and shared) and two charging strategies (unidirectional and V2G). This paper further investigates the impact of the following factors:
\begin{itemize}
    \item \ac{BAU} schedules (cost minimised, \ac{MEF} minimized and unoptimized)
    \item Lead time
    \item Duration of the requested flexibility window
\end{itemize}

\section{Models and Methods}
This section explains how different charging models are simulated based on real charging transactions to quantify the flexibility potential of \ac{EV} fleets in delivering the two congestion management products - redispatch and capacity limitation.\par

We consider a set $\mathcal{N}$ of charging transactions indexed by $n$, each requiring a total charge of $\bar{e}_n$ to be delivered between their arrival time ($t^a_n$) and departure time ($t^d_n$). Charging occurs in discrete time intervals ($\Delta t$), during which the charging power ($p_{n,t}$) of each EV is constant. The charging power $p_{n,t}$ is a continuous variable, constrained between a maximum power ($\bar{p}_n$) and a minimum power ($\underline{p}_n$). For bidirectional charging, $\underline{p}_n = -\bar{p}_n$; otherwise, $\underline{p}_n = 0$. $\mathcal{T}$ (indexed by $t$) represents the set of time intervals over which the optimization problem is solved. In contrast, $\mathcal{T}_f \subseteq \mathcal{T}$ denotes the subset of time intervals during which a specific congestion product is requested. \par
Each charging transaction is required to fulfil the energy volume derived from historical data, which may be less than the battery capacity (because the \ac{EV} was not empty at the start of the session or not full at the end). For bi-directional charging, the batteries' state of charge (SOC) cannot drop below their initial SOC, ensuring that \ac{EV} owners never find their vehicles with a lower SOC than at the start. Because \acp{EV} are restricted from charging beyond their historical charge volumes, resulting in a conservative estimate of feasible flexibility. Data originally recorded in seconds is resampled to 15-minute intervals, with arrival and departure times rounded to the nearest 15 minutes. Transactions that become infeasible after rounding due to insufficient connection time are excluded (less than 0.1\% of transactions). The maximum connection duration for an \ac{EV} is capped at 24 hours (96 time steps).

\subsection{\ac{BAU} dispatch strategies}
This paper considers three \ac{BAU} dispatch strategies: cost minimized, \ac{MEF} minimized and unoptimized. The \ac{BAU} optimized profile ($\tilde{p}^{*}_{n,t}$) is calculated by solving the following linear optimization problem.\par
Objective \eqref{obj:cost} minimizes the total cost of charging based on day-ahead prices ($\Pi^{\text{da}}_t$, \euro{}/kWh), \eqref{obj:mef} minimizes the marginal emissions ($\Pi^{\text{mef}}_t$, kgCO\textsubscript{2}/kwh) and \eqref{obj:unoptimized} makes sure all the \acp{EV} are charged as fast as possible (unoptimized or `dumb' charging). Further, all the \acp{EV} are constrained by \eqref{eq:energy_limit1}-\eqref{eq:power_limit2}, which holds the above assumptions.
\begin{subnumcases}{ \min_p f(p);\quad f(p) =}
    \sum_{t\in \mathcal{T}}\sum_{n\in \mathcal{N}}\Pi^{\text{da}}_t p_{n,t} \Delta t\label{obj:cost} \\
\sum_{t\in \mathcal{T}}\sum_{n\in \mathcal{N}}\Pi^{\text{mef}}_t p_{n,t} \Delta  )\label{obj:mef}\\
\sum_{t\in \mathcal{T}}\sum_{n\in \mathcal{N}}e_{n,t} \label{obj:unoptimized}
\end{subnumcases}
subject to:
\begin{align}
    &e_{n,t} = 0, && t\leq t_n^a,\:&&n\in\mathcal{N}
    \label{eq:energy_limit1}  \\
    &e_{n,t} = e_{n, t-1}+p_{n,t-1}\Delta t   ,&&t_n^a <t<t_n^d,\:&&n\in\mathcal{N}\label{eq:energy_change}\\
    &e_{n,t} = \overline{e}_n ,&&t\geq t_n^d  ,\:&&n\in\mathcal{N}\label{eq:energy_limit2}\\
    & p_{n,t}  = 0 ,&& t<t_n^a \lor t\geq t_n^d ,\:&&n\in\mathcal{N}\label{eq:power_limit1}\\
    & \underline{p}_{n}\leq  p_{n,t} \leq \overline{p}_{n} ,&& t_n^a\leq t < t_n^d ,\:&&n\in\mathcal{N}\label{eq:power_limit2} 
\end{align}

\begin{figure}
    \centering
    \includegraphics[width=0.95\linewidth]{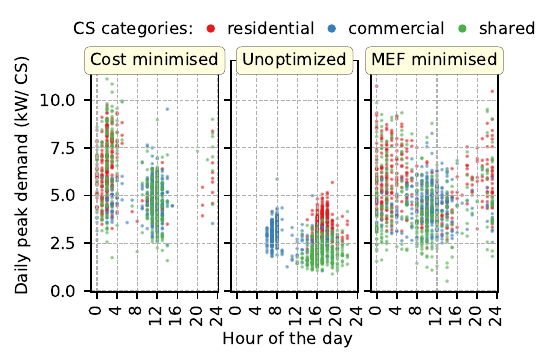}
    \caption{Relationship between daily peak power usage and the hour of the day across all days in the year 2023. Each point represents the peak power recorded at a specific hour for a particular \ac{BAU} schedule and \ac{CS} category.}
    \label{fig:hourly_peak_for_year}
\end{figure}

As illustrated in \figref{fig:hourly_peak_for_year}, the daily peak consumption occurs at different hours of the day for different \ac{BAU} schedules. Optimized profiles based on day-ahead prices and unoptimized profiles exhibit higher peaks than unoptimized charging. The magnitude of these peaks depends on the density of \ac{EV} arrivals at a given hour and the timing of lower electricity prices or \ac{MEF}, with the  \ac{MEF}-minimized profile showing a more diffuse pattern. Hence, due to these distinct variations, it is important to investigate the impact of different \ac{BAU} schedules on the delivery of congestion management products.
\begin{figure*}
    \centering
    \includegraphics[width=0.9\linewidth]{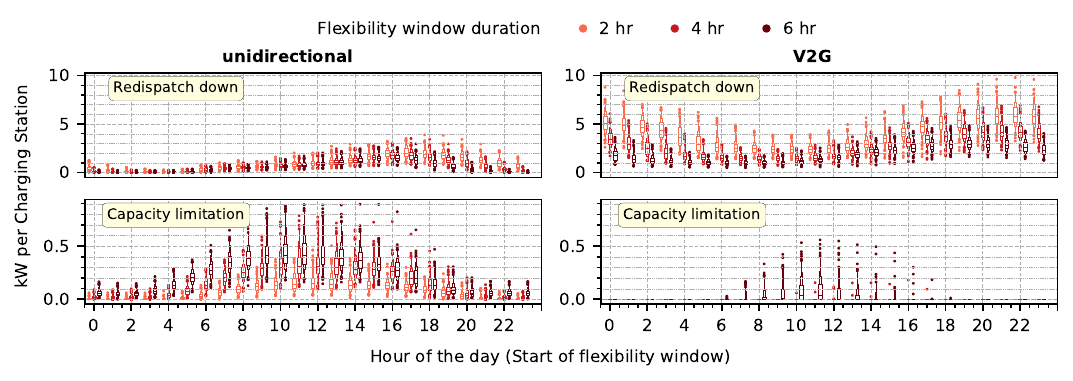} 
    \caption{Distribution of flexibility (redispatch and capacity limitation) for unoptimized \ac{BAU} schedule shown for different start of flexibility request windows of 1 hour. For the purpose of illustration, only residential charging stations are shown.}
    \label{fig:variation_window_duration}
\end{figure*}
\subsection{Flexibility products}
This paper analyses two congestion management products the aggregator must deliver during the requested flexibility window $\mathcal{T}_f$. The information about the delivery window ($\mathcal{T}_f$) is assumed to be known to the aggregator only $l$ hours (lead time) in advance. In our case study, we consider a single aggregator who also serves as the \ac{CPO}, responsible for scheduling the \acp{EV}. The optimized profiles are obtained by solving a linear programming problem described below.
\subsubsection{Optimal downward redispatch }
The optimal downward redispatch strategy aims to maximize the downward deviation between the aggregated power profile and the optimized \ac{BAU} profile ($\tilde{p}^{*}_{n,t}$). We assume perfect knowledge of the arrival and departure times of all \acp{EV}. The optimal downward redispatch ($c^r$) is calculated by maximizing \eqref{eq:objective_redispatch}, where the secondary objective based on the desired \ac{BAU} case is multiplied with an auxiliary parameter ($\epsilon$). Based on sensitivity analysis, the value of $\epsilon$ was fixed to $1*10^{-6}$.
\begin{equation}
    \max_{c^r, p, e}\: c^r + \epsilon f(p)
    \label{eq:objective_redispatch}
\end{equation}
subject to:
\begin{align}
   &\text{constraints \eqref{eq:energy_limit1}-\eqref{eq:power_limit2}} \nonumber \\ 
&\sum_{n\in\mathcal{N}}p_{n,t} \leq \tilde{p}^{*}_{n,t} - c^r, &&\forall t\in\mathcal{T}_f \label{eq:constraint_redispatch}\\
&p_{n,t} = \tilde{p}^{*}_{n,t}, &&\forall t< \min(\mathcal{T}_f)- l/\Delta t, \:\forall n \in \mathcal{N}
\end{align}
\subsubsection{Optimal capacity limitation}
Optimal capacity limitation strategies dispatch \acp{EV} to minimize the maximum aggregate power during a given flexibility request window. Like the optimal redispatch policy, this strategy also assumes perfect knowledge of all \ac{EV} arrival and departure times. The optimal capacity limitation ($c^l$) is also calculated using a linear optimization problem, where it is assumed that $c^l \ge 0$. Similar to the redispatch product, this can be solved for different \ac{BAU} profiles.
\begin{equation}
    \min_{c^l, p, e}\: c^l + \epsilon \left(f(p)\right)
    \label{eq:objective_cap_lim}
\end{equation}
subject to:
\begin{align}
   &\text{constraints \eqref{eq:energy_limit1}-\eqref{eq:power_limit2}} \nonumber \\ 
   & c^l \ge 0 &&\\
&\sum_{n\in\mathcal{N}}p_{n,t} \leq  c^r&&,t\in\mathcal{T}_f \label{eq:constraint_cap_lim}\\
&p_{n,t} = \tilde{p}^{*}_{n,t} &&, \forall n \in \mathcal{N} \:\: t\leq \min(\mathcal{T}_f)- l/\Delta t
\end{align}

\section{Results}
\subsection{Sampling of real \ac{EV} transactions}
The analysis considers over 350,000 real charging transactions in the Netherlands during 2023. In the selected data, there are 668 unique charging stations, with 313 in the residential category, 254 in the commercial category, and 101 in the shared category. The categorization of the \acp{CS} was done as explained in \cite{panda2024quantifying}. The transactions are distributed among three categories of \acp{CS} - residential(57.6 \%), commercial (31 \%), shared (11.4 \%). Each \ac{CS} has two connectors that can be connected to separate \acp{EV}.\par
In the analysis, each day of the year was taken as an independent sample. All transactions with arrivals on the sample date and the previous day are considered to simulate the \emph{sampled date}. The optimization horizon spans one day before and one day after the sample day, ensuring that all transactions can be fully charged. In the analysis, day-ahead energy prices for 2023 were used from the ENTSO-E Transparency platform\footnote{\url{https://newtransparency.entsoe.eu/}}. The \ac{MEF} data was obtained from the authors of~\cite{alikhani2023marginal}.
\subsection{Variation of flexibility window duration}
With the increase in the length of the flexibility window, the ability to minimise the power decreases due to the reduction in charging flexibility for both the products, as illustrated in \figref{fig:variation_window_duration} for residential \acp{CS} with an unoptimized \ac{BAU} schedule. The available redispatch flexibility increases with the arrival density of \acp{EV}. However, the minimum required capacity peaks (so the ability to constrain capacity drops) around mid-day when short charging sessions dominate. The overall magnitude of tradable flexibility sharply increases for \ac{V2G}, while still reducing gradually with increase in the flexibility request duration.
\subsection{Variation in lead time}
\begin{figure*}
    \centering
    \includegraphics[width=0.9\linewidth]{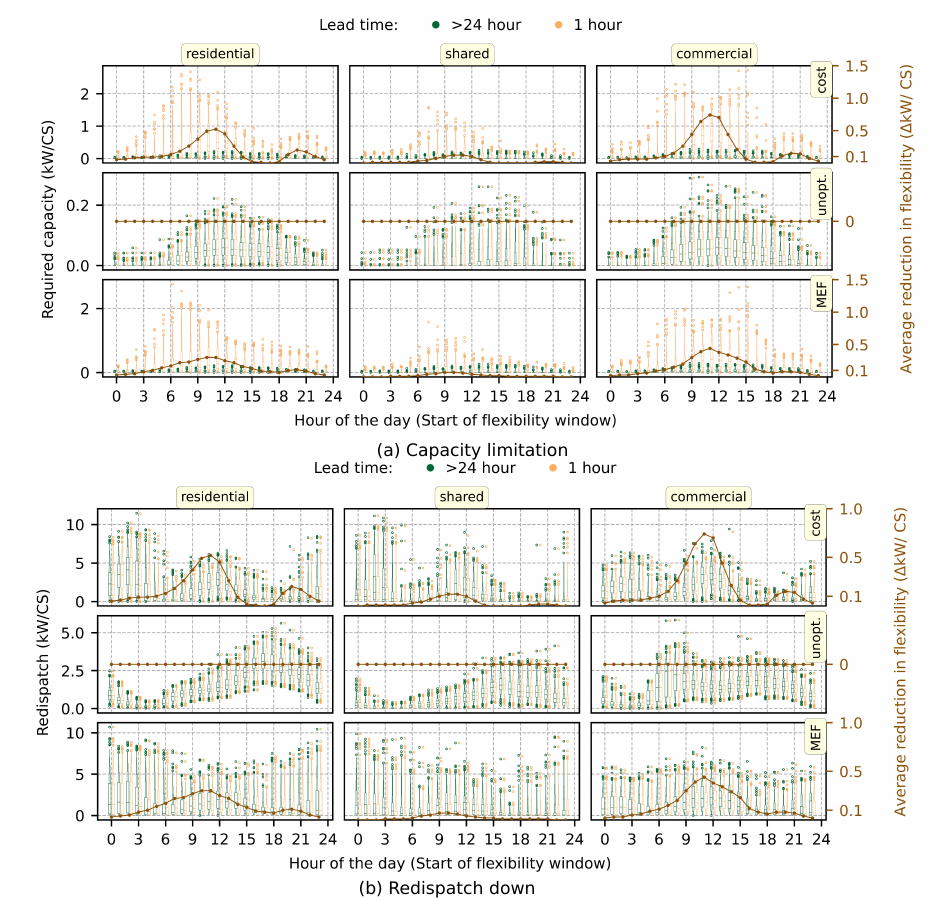}
    \caption{Impact of different lead times compared across different \ac{CS} categories and \ac{BAU} schedules. The difference between the average flexibility magnitude for the two lead times has been plotted using the secondary axis. Flexible windows with different start times and a duration of 1 hour are chosen.}
    \label{fig:lead_time_variation}
\end{figure*}
\begin{figure*}
    \centering
    \includegraphics[width=0.9\linewidth]{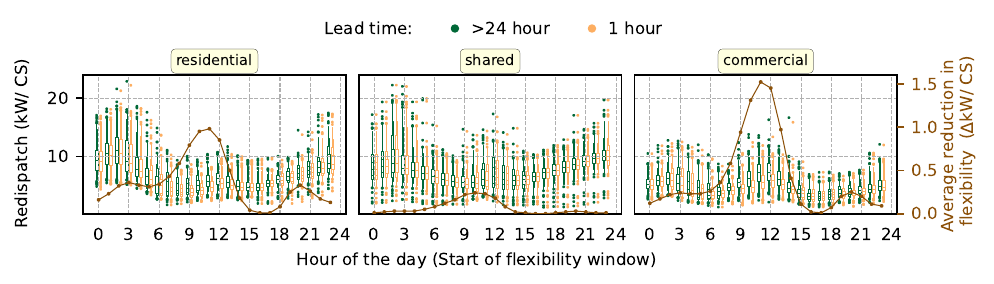}
    \caption{Effect of different lead times on the flexibility for redispatch with \ac{V2G}. For the purpose of this illustration, only the case with cost minimized \ac{BAU} schedule is shown for different categories of \ac{CS} for a flexibility window of one hour.}
    \label{fig:lead_time_v2g}
\end{figure*}
Two different lead times - day-ahead activation and activation 1 hour ahead of expected delivery are chosen as two possible extreme cases, which also illustrate the trend for in-between cases. \figref{fig:lead_time_variation}-(a) illustrates the reduction in the \ac{EV} fleets' ability to limit their consumption during the requested flexibility windows of one hour for the two chosen lead times. The results are compared across the three charging station categories for different \ac{BAU} schedules, considering only unidirectional smart charging. Tradable flexibility significantly decreases with shorter lead times for all \ac{BAU} schedules, except in the unoptimized case. This is because unoptimized charging is a greedy strategy that is unaffected by lead time. Higher average reductions in tradable flexibility coincide with lower average day-ahead and \ac{MEF} prices. During such a period, limiting power depends largely on schedule adaption before the delivery. Hence, with shorter lead time, the \ac{EV} schedules are less able to adapt because they were originally scheduled to charge during lower-cost periods

\figref{fig:lead_time_variation}-(b) illustrates similar results, but for the redispatch product. As a general trend, the magnitude of redispatch capability depends on the density of \ac{EV} arrivals and the timing of lower energy/ emission costs. It further scales with the maximum charging power, occupancy of \acp{CS} and the energy needs, which are highly dependent on the charging station's category. Notably, the \emph{reduction} in tradable flexibility as a result of a reduction in lead time is the same as for the capacity limitation product. This can be understood because the reduction depends only on the ability to adjust schedules between activation and delivery, which is the same irrespective of the product.\par
Using \ac{V2G}, the \acp{EV} can more flexibly adapt their schedule between activation and delivery. This is due to the possibility of aggregate power reduction by discharging some \acp{EV} while compensating for other \acp{EV} that must be charged. This trend is clearly illustrated in \figref{fig:lead_time_v2g}, which plots the tradable redispatch flexibility for different one-hour flexibility windows across \ac{CS} categories for the cost-minimized \ac{BAU} schedule. Similar trends are also observed for other \ac{BAU} schedules. The redispatch flexibility increases greatly with \ac{V2G}. At the same time, although its \emph{reduction} due to a shorter notice period increases more compared to the unidirectional case, the overall magnitude is still higher than unidirectional smart charging. Capacity limitation results are not shown because the flexibility of \ac{V2G} is such that a capacity reduction to $0$kW is feasible for nearly all scenarios. 

\subsection{Implication of lead time on average charging costs}
\begin{figure}
    \centering
    \includegraphics[width=0.9\linewidth]{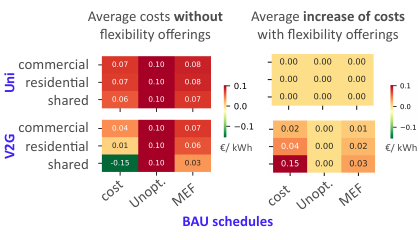}
    \caption{Heatmap illustrating the hourly averaged energy costs under different \ac{BAU} schedules (left). The right heatmap shows the corresponding increase in average costs after adapting schedules to deliver flexibility for congestion management (capacity limitation product).}
    \label{fig:cost_var}
\end{figure}

When opting to adjust charging to deliver congestion management services, the aggregator's energy costs can increase. To illustrate the impact of the \ac{BAU} scheduling strategy and the impact of delivering services,  \figref{fig:cost_var} shows the average cost of charging for 2023 and its change when flexibility is provided in the form of the capacity limitation. The energy costs are consistently lower for \ac{V2G} compared to the unidirectional charging when all other parameters are kept the same. However, this does not consider the required investments in hardware (\ac{EV}, \ac{CS}), extra wear and tear and control system requirements. Further, we show the average cost increase when the \ac{BAU} schedules are updated to deliver flexibility services for congestion management. For unidirectional charging, no cost increase is observed. However, there is an increase in costs under V2G charging depending on the \ac{BAU} schedule - largely offsetting the cost gains made in the no-flexibility scenario. 

\section{Conclusion}
\balance
In this paper, we investigated the effects of different lead times and window lengths on the magnitude of tradable flexibility for different products compared to three \ac{CS} categories and \ac{BAU} schedules. Shorter lead times reduce tradable flexibility in all cases except in the case of an unoptimized \ac{BAU} schedule, where charging as soon as connected (a greedy strategy) limits the charging adaptability. \ac{V2G} enables more flexible adaption of schedules, thereby reducing the effect of lead times on the delivery of flexibility. Reduction trends are also observed with the increased flexibility window duration, which decreases with the adoption of \ac{V2G}. This paper thus highlights the usefulness of \ac{V2G} when uncertainty regarding lead times cannot be completely removed. Further research regarding the cost and value of flexible products for different stakeholders can complement the presented research.
\section*{Acknowledgment}
The authors thank P. Alikhani of Copernicus Institute of Sustainable Development, Utrecht University, The Netherlands, for supplying us with the marginal emission factor data used in the paper. The authors also acknowledge the use of computational resources of Snellius - the National Supercomputer provided by Surf, the ICT cooperative of Dutch education and research institutions.
\bibliographystyle{IEEEtran}
\bibliography{ref}
\end{document}